\documentclass[a4paper]{jpconf}

\def\CC{{C\nolinebreak[4]\hspace{-.05em}\raisebox{.4ex}{\tiny\bf ++}}}

\begin{document}

\title{GPU molecular dynamics: Algorithms and performance}

\author{D C Rapaport}

\address{Department of Physics, Bar-Ilan University, Ramat-Gan 52900, Israel}

\ead{rapaport@mail.biu.ac.il}

%% 02/20

\begin{abstract}

A previous study of MD algorithms designed for GPU use is extended to cover more 
recent developments in GPU architecture. Algorithm modifications are described, 
together with extensions to more complex systems. New measurements include the effects 
of increased parallelism on GPU performance, as well as comparisons with 
multiple-core CPUs using multitasking based on CPU threads and message passing. 
The results show that the GPU retains a significant performance advantage.

\end{abstract}

\section{Introduction}

Improved computer performance is increasingly dependent on parallelism, a 
consequence of clock limits for current processor technology. Automating simpler 
forms of software parallelization is within the capability of modern compilers, 
but manual software redesign is required whenever the algorithm logic is 
incompatible with the processor architecture. This was true for the earliest 
vector and parallel supercomputers, and is equally so for modern hardware in which 
the GPU, or graphic processing unit, a device whose name reflects its original 
purpose, is the key component.

While the physical phenomena modeled by MD (molecular dynamics) simulation 
\cite{mdbk} have a major parallel component, in which particles (atoms or 
molecules) independently `evaluate' the combined forces exerted on them by their 
surroundings, MD computations are traditionally implemented in serial form, 
namely, at each time step there are processing loops that iterate over all 
particles or particle-pairs. The availability of large-scale data parallelism 
eliminates the need for most of these explicit loops. As examples, the overall 
force computation for each particle can be carried out completely independently 
and in parallel (provided Newton's third law is not invoked), while the trivial 
sum to evaluate the total energy is inherently serial and therefore requires the 
use of efficient, but more complex, parallel reduction techniques.

The present paper considers MD algorithm implementation on a more advanced GPU 
than considered previously \cite{rap11a}. Measured speedups are due not only to 
the substantially increased hardware parallelism but also to architectural 
enhancements. In addition to the comparison between GPU performance and a 
single-threaded CPU application, comparisons have been made with alternative 
approaches employing multicore CPU parallelism based on CPU (or Unix) threads 
and message-passing. Factors associated with the algorithms and their 
implementations that influence GPU performance are also investigated.

\section{Background and methodology}

\subsection{GPU hardware}

The GPU is a highly parallel coprocessor, with its own private memory, that 
operates separately from the CPU but under its control. GPUs provide extremely 
fine-grained data parallelism, unlike a typical CPU with at most a few 
processing cores. Data can be exchanged between CPU and GPU, although for 
efficiency this should be kept to a minimum. GPU hardware, and the associated 
software, are in a state of continuous development, but have already reached a 
level of maturity where GPUs are the key components in many of the most powerful 
supercomputers currently available, while at the same time also available 
packaged as affordable commodity products, principally plugin graphics cards and 
chipsets. This suggests that GPU-based computing is destined for a long life, 
justifying the investment of effort in designing algorithms tuned for optimal 
GPU performance (product lifetime is a factor in determining the overall 
cost-effectiveness of novel hardware for general use). Current GPU complexity 
tends to require software design by trial and error, with improvements over 
successive hardware generations providing more flexibility that can simplify the 
task.

GPUs are assembled from a set of multiple-core processors and eschew many of the 
more complex features of conventional CPUs; the performance gains are achieved 
mainly by parallelism rather than the more sophisticated CPU instruction 
handling. To use NVIDIA nomenclature \cite{cudaprog}, a GPU device consists of 
multiple streaming multiprocessors (SMs), each consisting of (typically) 128 
cores. A fully-configured GPU contains up to several thousand cores. Each core 
executes a single instruction thread, with threads grouped into blocks that 
execute as multiple 32-thread warps on each SM. While the threads in each block 
can be synchronized and can exchange data through shared memory, separate blocks 
are processed independently and there is no direct communication between them, 
with block synchronization typically occuring at kernel (defined below) 
completion. Other, more subtle, details associated with thread scheduling are of 
more specialized interest \cite{cudaprog,pasctun}.

The GPU includes onboard global memory; since the access latency is relatively 
long, measured in hundreds of clock cycles, the ability to coalesce memory 
requests improves read/write speeds substantially. In its simplest form, 
coalescence requires that blocks of threads access aligned, consecutive memory 
locations. Memory latency can also be hidden when there are multiple thread 
warps awaiting processing. Since blocks access global memory completely 
independently, synchronization between blocks within a kernel is nontrivial and 
requires special effort. Multiple caches can improve noncoalesced memory usage, 
a feature that varies between GPU generations. Global memory is also accessible 
by the host. Threads within the same block can also access much faster 
read/write shared memory that enables exchange of data (there are possible bank 
conflicts that can slow access, as well as the need for thread synchronization 
to prevent indeterminate race conditions). There is also a fast read-only 
constant memory that the host can write into. Finally, there is a set of local 
registers private to each thread. Further details of this complex memory 
scenario appear in \cite{cudaprog,tespap}.

Function (or subroutine) calls are executed as GPU kernels. Each kernel call 
from the CPU specifies the size of the thread block and the total number of 
blocks. Multiple thread blocks run concurrently in an unspecified order 
determined by a scheduler. Explicit thread synchronization is available to 
ensure that threads reach a particular way-point in their execution, e.g., prior 
to exchanging data, although care is required to avoid deadlocks and not degrade 
performance. Software development typically involves CUDA C (or \CC), a language 
extension that greatly simplifies GPU programming.

\subsection{MD algorithms}

Algorithms designed to utilize massive parallelism introduce novel issues. Some 
computations are readily converted from serial to parallel form, especially 
when independent threads can replace iterated operations that have no mutual 
effect. Others require cooperation between threads because data must be shared, 
such as the reduction-type operation involved in summing an array, and in more 
complex tasks such as sorting; these require completely different techniques 
that are inefficient serially but optimal when parallelized. Some changes can be 
handled using standardized approaches (e.g., reduction and sorting), while 
others need special treatment, notably the MD tabulation of atom neighbors using 
the neighbor matrix (below), that replaces the more familiar neighbor list due 
to its unsuitability for parallel use.

The three kinds of MD programs \cite{mdbk} that are considered here, aside from 
the standard serial version, are those based on job subdivision using (a) 
multiple CPU threads, (b) coarse-grained parallelism in which several CPU tasks, 
with message-passing communication, are responsible for subregions of the 
system, and (c) fine-grained parallelism with one GPU thread per atom. Each 
entails changes from the serial approach; in terms of effort to adapt the 
programs, CPU threads are the easiest, message-passing requires extra 
bookkeeping, while the GPU-based approach requires major reformulation 
resembling that used in vectorization \cite{rap06a}.

The GPU algorithm uses a matrix array for cell occupancy whose filling and 
occupancy counting require `atomic' update operations (in this context `atomic' 
means an uninterruptible sequence of operations that are guaranteed exclusive 
memory access). This matrix is then accessed in the orthogonal direction to 
obtain the layers that contain the sets of interaction neighbors of each atom; 
here there are also different ways of organizing the computation, with the 
method described here being efficient and fully scalable.

\subsection{Neighbor matrix construction and usage}

The algorithm details, with an emphasis on the differences between the CPU and 
GPU approaches, are described in \cite{rap11a}. A brief summary of the GPU 
version of the method follows to allow subsequent performance measurements to 
reference the steps. Practically all the computation is carried out on the GPU, 
with minimal CPU involvement. Additional details of the MD computation are 
covered in \cite{rap11a} and more general considerations appear in \cite{mdbk}.

In all cases, loops over atoms in the serial CPU program, or the outer loops 
when nested, are replaced by GPU threads, one per atom. Maximal parallelism is 
achieved, with performance scaling limited only by hardware capability, 
primarily the GPU core count and clock speed, and the memory bandwidth.
Neighbor matrix construction is carried out at intervals of several time steps:

(N1) Assign atoms to cells, based on position; atom $i$ is in cell $c_i$, with 
multiple atoms allowed per cell.

(N2) Assign atoms in each cell to layers, where layer $l$ includes the $l$th 
members of each cell; atom $i$ is in layer $l_i$. Each cell $c$ requires a cell 
occupancy counter $k_c$, and due to multiple occupancy, incrementing $k_{c_i}$ 
requires an `atomic' operation. 

(N3) Determine the number of layers required, $N_l$, by finding the maximum
of $k_c$ using a reduction operation.

(N4) Build the cell-layer occupancy matrix $H$ by setting $H_{c_i, l_i} = i$ for 
each atom $i$; the row and column indices of $H_{c, l}$, specify the cell ($c \le 
N_c$) and layer ($l \le N_l$).

(N5) Construct the neighbor matrix $W$; for each atom $i$ there are two nested 
loops to access the neighbors $i' = H_{c, l}$, first over the neighboring cells 
$c$ of $c_i$, and then over layers $l \le k_c$ . The column indices $i$ of 
$W_{m, i}$ correspond to atoms; each row $m$ specifies the atoms' $m$th 
neighbors (unordered). The count of $i$'s neighbors is accumulated in $m_i$.

Force evaluation, at each time step:

(F1) For each atom $i$ accumulate the total force $\vec{f}_i$ and (optionally) 
interaction energy $u_i$ by considering the subset of atoms $i' = W_{m, i}$ for 
$m < m_i$ that lie within interaction range.

(F2) Sum the individual $u_{i}$ to obtain the total interaction energy $U$ 
(actually $2 U$) using a reduction operation (optional).

Comments on the computations follow, including ways that the GPU architecture 
affects the algorithm.

There can be a performance penalty for `atomic' GPU operations; this depends on 
the GPU architecture. In \cite{rap11a} better performance was obtained when the 
$l_i$ evaluation (N2) was carried out on the CPU, even after allowing for the 
additional data transfers. With more recent GPUs this is no longer the case. 
Note that situations requiring `atomic' operations also lead to irreproducible 
floating-point results since the order of layer occupants in each cell is 
indeterminate; changing the order in which atoms are processed can alter the 
lowest-order mantissa bits because the arithmetic is nonassociative, but for MD 
this sensitivity is masked by the inherently chaotic nature of the atomic 
trajectories \cite{mdbk}.

Efficient evaluation of global quantities on the GPU requires (nontrivial) 
parallel reduction operations; examples include the total interaction and 
kinetic energies, and the maximal atom velocity needed to determine when the 
neighbor matrix needs rebuilding. Reduction operations rely on efficient use of 
shared memory and follow a standard hierarchical pattern that maximizes the work 
carried out in parallel \cite{harred}. In (N3) and (F2) the reductions are 
carried out on the GPU, producing a single result per thread block, with a minor 
completion step on the CPU after a small data transfer.

The final reductions, involving one result per thread block, can also be carried 
out on the GPU. Since the number of thread blocks is much less than the number 
of atoms, by two orders of magnitude, and the final result is usually required 
by the CPU, the performance benefit of the extra GPU work is limited, but it 
does allow greater task separation between GPU and CPU. The GPU implementation 
\cite{cudaprog} uses an `atomic'ally updated counter in global memory to record 
the number of blocks that complete their work (since thread blocks are 
independent this is the most a block can know about mutual progress) as well as 
a `memory fence' (lockout) function to ensure that write operations are 
completed before proceeding. The block that finds itself to be the last to 
finish can then finalize the reduction task, using its threads to process the 
data output by all blocks (itself included) to global memory, yielding a single 
result.

The layout of $W$, assuming the matrix to be stored by rows, allows the 
identities of all $m$th neighbors to occupy successive memory locations; this 
permits coalesced access by threads processing individual atoms in (F1). Atom 
pairs appear twice in $W$ and are considered twice in (F1) because Newton's 
third law is not invoked; instead of using each atom pair $(i, i')$ to update 
the (equal but opposite) forces acting on both atoms, $\vec{f}_i$ and 
$\vec{f}_{i'}$, where atoms $i$ are accessed sequentially but atoms $i'$ 
essentially at random, only $\vec{f}_i$ is updated. This allows more efficient 
coalesced memory access for atoms $i$ that compensates for the extra computation.

\subsection{Software design}

A number of considerations influenced the overall software design.

(1) The variables describing the state of each atom, namely its position, 
velocity and acceleration, are three-component vectors. Representing them on the 
GPU as {\tt float4} quantities allows efficient memory access, more than 
justifying the extra storage; the unused fourth component can hold other 
information, such as the interaction energy during the force calculation. Use of 
an array of structures that combine different kinds of state data for each atom, 
such as position and velocity, is a `recommended' form of data organization; 
unfortunately, it is incompatible with efficient GPU operation since it inhibits 
coalesced memory access and degrades performance.

(2) Memory allocation on the GPU is requested by CPU function calls. The GPU 
memory pointers returned must later be supplied to the GPU. This could be done 
by including them as arguments to GPU kernel calls. However, since the 
computations involve several arrays as well as other parameters, all this 
unvarying information is placed in a small data structure and copied to a region 
of GPU storage called the constant memory where it can be efficiently accessed 
by all threads; this eliminates the need for long argument lists that would 
otherwise accompany each kernel call.

(3) Essentially all computation is done on the GPU to avoid relatively slow CPU-GPU 
data transfers. After constructing the initial state on the CPU, almost all the 
state data exists only on the GPU. Results of measurements are returned to the 
CPU when required. If visualization is employed, atom coordinates must be made 
available by the GPU for updating the imagery; if a single GPU handles both 
computation and graphics, even this data transfer can be avoided. Checkpoint and 
restart would also require transfers of most of the state data.

(4) An exception to using the GPU for everything is the infrequent data 
rearrangement (sorting) required to reorder atoms for efficient memory access. 
The problem is to pack the nonzero elements of $H$ into a vector $s$ of length 
$N_a$ that specifies the reordered atom sequence (determined by cell 
membership); though trivial when carried out serially, the efficient parallel 
version is more complicated. Here, reordering involves copying the most recent 
version of $H$ to the CPU; this is used to construct $s$ which is returned to 
the GPU and used to reorder atom coordinates and velocities (using a temporary 
buffer, with caching aided by the fact that data is already partially ordered), 
following which the matrices $H$ and $W$ are rebuilt. These data transfers have 
negligible performance impact. The GPU alternative would be to apply a parallel 
prefix-scan \cite{har07} to the set of $k_c$ that specify the cell occupancy; 
each element of the result is a cumulative sum of $k_c$ for the preceding cells. 
Cells are then processed in parallel, since each thread knows where to place its 
cell's atom identities in $s$. In general, small serial tasks that require 
excessive effort to implement on the GPU are worth converting only if there is a 
clear need to avoid data transfers with the host, or if suitable, well-tuned 
library code is available. A similar compromise was adopted with the reduction 
operations used for energy (etc.) evaluation (see above).

(5) There is an advantage to having a single program source that, during the 
compilation phase, generates code for either just the CPU or the combined 
GPU/CPU environment, rather than having to maintain separate versions; 
conditional statements in the source file(s) control this process. The NVIDIA 
{\tt nvcc} compiler \cite{cudaprog} deals with all the GPU kernel code and then 
automatically calls the {\tt gcc} C compiler to process the CPU code and the 
linking; for the CPU version of the program, only the C compiler is used. 
Debugging is simpler on the CPU, so that issues not specific to the GPU can be 
resolved simultaneously in both versions of the code. This approach assumes that 
the same algorithm is being used in both cases and the changes are principally 
the replacement of loops by parallel threads, as well as standardized reduction 
procedures, which indeed cover the most frequent changes when converting CPU 
code for GPU use.

\section{Performance measurements}

\subsection{Hardware and software configuration}

The NVIDIA GPU considered here is the mobile version of the P4000, based on the 
(recent, but not latest) Pascal architecture \cite{tespap}. Detailed GPU design 
is subject to frequent change, with some changes affecting performance 
significantly, and others less so; the consequences are not always apparent from 
the hardware specifications. The principal feature characterizing a GPU is the 
core count, reflecting its parallel capability. Given the limits to Moore's law, 
increasing core count is the main route to faster performance; the P4000 has 
1792 cores. Processor clock speed is also important; here it is 1227MHz, 
however, what the GPU is capable of doing in a single cycle (e.g., multiple 
operations per thread) varies with model. Memory access is complicated 
\cite{cudaprog,pasctun,cudaprac}, but the overall bandwidth is important; here 
the value is 192 GB/s. There are numerous other contributing factors, but the 
nominal peak rating is 4398 s.p. (single precision) GFlop/s. While these numbers 
are purely theoretical, and unachievable, the difference is reflected in the 
actual performance. By comparison, the earlier \cite{rap11a} FX770M  had a very 
modest 32 cores, 500MHz clock, 26 GB/s bandwidth, and a peak 80 GFlop/s, so a 
considerable speedup can be anticipated.

Several aspects of performance will be analyzed. Comparisons of GPU speed relative 
to a single CPU core are considered first. These are followed by tests comparing 
the increased CPU capability made available by using multiple CPU cores in two 
different ways. Factors contributing to GPU performance are considered; these 
can sometimes reveal aspects of behavior not obvious from the published device 
specifications.

The system parameters used for the tests strongly affect the results; they are 
the same as in \cite{rap11a} and are as follows (all in reduced MD units), 
unless otherwise specified: The soft-sphere system, denoted by SP, has density 
$\rho = 0.8$, temperature $T = 1$, and interaction cutoff $r_c = 2^{1/6} = 
1.122$. For the LJ system the values are $\rho = 0.38$, $T = 1.2$ and $r_c = 
2.5$. In both cases particle reordering occurs every 100 time steps, the shell 
thickness used for neighbor matrix is $\delta = 0.6$ (in \cite{rap11a} 0.8 was 
used for LJ), and the integration time step is 0.005. The initial state consists 
of a cubic grid of atoms of size $N_a = {N_e}^3$ that are assigned random 
velocities. Runs are of length 5000 time steps.

GPU measurements were carried out on a Dell Precision 7720 notebook workstation 
(Intel 4-core Xeon E3-1505Mv6 CPU) running Ubuntu Linux; GPU development is 
based on CUDA version 8. The other computer referenced in this work (for the 
multicore CPU comparisons) is an HP Z820 dual-CPU workstation (two 4-core Intel 
Xeon E5-2637v2 CPUs) running Centos Linux (with d.p. computation that is only 
slightly slower than s.p.); compilation uses the {\tt gcc} compiler, run at 
optimization level {\tt O3}; the CPU comparisons use the additional cores of 
this slightly faster processor pair.

\subsection{GPU tests}

A series of GPU speed measurements for the SP and LJ systems over a range of 
sizes, extending beyond those considered in \cite{rap11a}, appear in Table 
\ref{tab1}. These results are for the most efficient version of the algorithm, 
run under optimal conditions; factors determining these conditions are discussed 
later. 

\begin{table}
\caption{\label{tab1}Size dependence for soft sphere (SP) and Lennard-Jones (LJ) 
systems; the numbers of atoms ($N_a = {N_e}^3$) and the times per atom-step ($t$, 
in $\mu s$) are shown.}
\begin{center}
\begin{tabular}{rrrll}
\br
      & $N_e$ &    $N_a$ &  $t_{SP}$ &  $t_{LJ}$  \\
\mr
 GPU  &   32 &	  32768	 &  0.00546  &  0.01152   \\
      &   64 &   262144  &  0.00363  &  0.00850   \\
      &   96 &   884736  &  0.00356  &  0.00881   \\
      &  128 &  2097152  &  0.00373  &  0.00925   \\
      &  160 &  4096000  &  0.00386  &  0.00898   \\
      &  192 &  7077888  &  0.00386  &  0.00919   \\
      &  224 & 11239424  &  0.00387  &            \\
      &  256 & 16777216  &  0.00386  &            \\
 CPU  &   48 &   110592  &  0.134    &  0.558     \\
\br
\end{tabular}
\end{center}
\end{table}

Over most of the size range, excluding the smallest system that does not allow 
the GPU to reach full capacity, there is no systematic size dependence (cell 
size has a small effect on the results). Results for the best CPU version of the 
MD algorithm on a single CPU core are included for comparison. The P4000 GPU is 
seen to provide speedups relative to the CPU of 34.7x and 60.7x for the SP and 
LJ cases. There is also a 20-fold speedup over the original FX770, that lies 
between the increased 8x memory bandwidth and 50x core count with doubled clock 
speed. Running the layer-based algorithm on the CPU halves the performance 
(allowing the dubious claim of a further 2x GPU speedup). 

\begin{table}
\caption{\label{tab2}Percentage of time spent in different parts of the MD 
calculation.}
\begin{center}
\begin{tabular}{lrr}
\br
  Task  	&  SP     &  LJ     \\
\mr
 nebr matrix:   &  44.0   &  58.1   \\
      forces:   &  26.8   &  29.8   \\
 integration:   &  20.1   &   8.4   \\
      energy:   &   5.9   &   2.5   \\  
 layer cells:   &   1.7   &   0.7   \\
     reorder:   &   1.4   &   0.4   \\
\br
\end{tabular}
\end{center}
\end{table}

Knowing where the computational effort is expended helps focus optimization 
attempts. Table \ref{tab2} shows the time fractions devoted to the steps of the 
calculation; the results depend on the GPU model and the system state. Work 
associated with neighbor matrix construction and force evaluation accounts for 
the majority of the processing time, with the former, manifestly unsuited for 
the GPU, dominating. The neighbor matrix is typically refreshed approximately 
every 10 steps (see below). On the CPU the situation is very different, and the 
respective time fractions for SP(LJ) are 0.26(0.22) and 0.63(0.76). Another 
consideration is increased storage of the layer approach, although here this is 
not an issue; storage per atom for SP(LJ) is 224(474) bytes, and the largest 
runs required a total of 3-4 GBytes out of the 8 GBytes installed.

\subsection{Multiple-core CPU comparisons}

Since modern CPUs incorporate multiple cores, comparisons of GPU and CPU 
performance should include parallel tests utilizing the available cores. The two 
ways of decomposing a computation to use one or both CPUs in a single processing 
node and the multicore CPUs themselves employ (a) CPU (or Unix) threads (no 
relation to GPU threads) that access common data, or (b) multiple CPU processes 
that communicate only via OpenMPI message-passing with each maintaining its own 
data storage, the latter directly transferable to multiple nodes connected over 
a network. In either case the MD software requires adaptation, although 
considerably less in the former; the details appear in \cite{mdbk}. Table 
\ref{tab3} summarizes tests run on the dual-CPU workstation that allowed up to 8 
threads and processes.  

\begin{table}
\caption{\label{tab3}Comparison of parallel MD approaches for the SP fluid based 
on CPU threads and MPI, with different thread and process counts ($n$); times 
per atom-step ($t$, in $\mu s$), speedup relative to serial code, and efficiency 
relative to the ideal case (no loss due to parallelism) are shown.}
\begin{center}
\begin{tabular}{lrcccc}
\br
        & $n$  &   $t$      & Speedup &  Eff.   \\
\mr
 1-core &      &   0.103    &         &         \\
  CPU   &   2  &   0.065    & 1.58x   &  79\%   \\
        &   4  &   0.041    & 2.51x   &  63\%   \\
        &   6  &   0.033    & 3.12x   &  52\%   \\
        &   8  &   0.030    & 3.43x   &  43\%   \\
  MPI   &   2  &   0.057    & 1.81x   &  91\%   \\
        &   4  &   0.030    & 3.43x   &  86\%   \\
        &   8  &   0.022    & 4.68x   &  59\%   \\
\br
\end{tabular}
\end{center}
\end{table}

The observed diminishing returns are due to the extra communication and 
computation needed to support increased parallelism. The GPU maintains its 
dominance, delivering 5x the speed of the dual-CPU machine; the corresponding LJ 
speedup (not shown) is 9x. (These results are unrelated to measurements that use 
multiple networked nodes where near-linear scaling is readily achieved.)

\section{Discussion}

\subsection{Factors influencing GPU performance}

Complex GPU architectures often allow a choice of algorithms, some with major 
impact. The results of the previous section were obtained after investigating 
different aspects of the algorithms. Although exhaustive testing was not carried 
out, possible dependencies were examined, some of which are discussed here.

(1) Reordering: For short-ranged interactions, caching can improve performance 
if atoms are indexed so that data for spatial neighbors are localized in only a 
few regions of GPU memory. This can be accomplished merely by rearranging the 
storage order of the atoms to correspond to the cells they occupy; atoms in the 
same cell need not be ordered. The processing speed drops gradually as the 
interval between reorder operations is increased. Reordering every 100 steps is 
about optimal, and the speed is 1.15x (SP, $N_e = 96$) slower when the interval 
is increased to 500 steps. Absent any sorting, the slowdown reaches 4.5x, a 
major performance loss that exceeds the earlier 2.5x \cite{rap11a}. Note that 
for measurements that involve the identities of individual atoms, e.g., 
diffusion, each must be assigned a permanent serial number since storage 
locations will change.

(2) Thread block size: In the GPU implementation, major loops are replaced by 
threads that are grouped into blocks. Multiple blocks are processed at the same 
time, allowing memory access latency to be hidden, but this is limited by 
model-dependent GPU resources, including any shared memory required by each 
block and the local (register) storage per thread. For the P4000 there is little 
sensitivity (3\%) to block size over the range 64-1024, so the value 256 was 
used; this is unlikely to be true in general.

(3) Texture cache: In tests with earlier GPUs the use of the texture cache for 
reading atom data from global memory produced a substantial (1.8x) speedup. This 
feature does not influence P4000 performance, although it does affect the 
somewhat older workstation Tesla K20C, and is likely due to a cache redesign for 
improving memory access \cite{pasctun}.

(4) Periodic boundaries: The extra computation involved is reduced \cite{rap11a} 
by using 6 bits in the neighbor matrix $W$ entries as flags describing the 
corrections for each pair ($\pm\, x$, etc.). Since most of the worked saved is in 
determining which (if any) corrections are required, 3 bits are sufficient, and 
the signs determined when needed. This is important in large (but not too large) 
systems where more than 26 bits of the 32-bit integers are needed for atom 
indexing.

(5) Neighbor shell width: A thicker shell ($\delta$) entails larger cells and 
hence more layers and neighbors; the compensation for increased storage is a 
reduced neighbor refresh rate.The optimal value (for each $\rho$ and $T$) must 
be determined empirically; here $\delta = 0.6$ is used. The largest SP(LJ) 
systems have, on average, 15(48) neighbors, 9(25) layers, and refreshing occurs 
every 10.2(9.8) time steps. Small irregular speed variations with system size 
can occur due to the integer number of cells in each grid direction.

(6) Matrix ordering: The access order of the $W$ matrix is a crucial factor. 
Here, transposing $W$ halves the performance. In the earlier work \cite{rap11a} 
the performance drop was just 1.1x, suggesting increased sensitivity to memory 
access issues (the ratio of cores/bandwidth of the GPUs also differs). The very 
existence of such strong sensitivity serves as a warning that care is required; 
even though $W$ is written only when the neighbors are refreshed, it is read 
every time step, and the matrix arrangement must facilitate coalesced access.

(7) Double-precision performance: On the P4000 the computations ran at about 
0.3x the speed, reflecting reduced hardware for d.p. computation. On the K20C 
the slowdown was just 0.65x. The proportion of work devoted to forces (typically 
60\%) dominates the d.p. computation, unlike the s.p. case (30\%); these 
comparisons do not reflect nominal GPU GFlop/s rates (d.p. vs s.p.) of 1/32x and 
1/3x. Use of d.p. has no obvious effect on the MD results: energy conservation 
and thermodynamic properties are unchanged; likewise dynamical quantities such 
as the velocity autocorrelation function of an SP fluid at high density ($\rho = 
1.0$, CPU tested), whose oscillations reveal that atoms are caged. 

(8) `Atomic' operations: Their efficiency improves with GPU generation. On the 
P4000 and the earlier K20C, their use in the cell-layer assignment (N2) had a 
negligible effect, but with the much older FX770M in \cite{rap11a} `atomic' 
operations resulted in 1.12x(1.04x) slowdowns for SP(LJ).

\subsection{Alternative approaches}

Two other methods are mentioned for comparison purposes.

(1) Cell-block method: An earlier approach to neighbor enumeration \cite{and08} 
associated GPU threads with cells instead of atoms, temporarily storing atom 
coordinates in shared memory. This was originally compared with the present 
layer method in \cite{rap11a} using the 32-core FX770M, and later confirmed 
using a faster 256-core Quadro K4000 GPU (unpublished). The latter measurements 
showed a layer speedup ranging from 1.2x for LJ with cutoff $r_c = 3.0$, 1.4x 
for $r_c = 2.5$, to 3.7x for SP, an improvement over the earlier values. Since 
the principal change to the method is how the neighbor matrix $W$ is 
constructed, it is interesting to compare the speedups for just this step. The 
results show corresponding speedup factors of 1.6x, 2.3x, and a substantial 
9.2x. Note that the overall time fraction for constructing $W$ increases as 
$r_c$ is reduced. 

(2) All-pairs force evaluation: Even more impressive performance comparisons 
favoring the GPU can be obtained using a naive MD approach that considers all 
possible atom pairs. The GPU is able to handle this very efficiently by using 
shared memory \cite{nyl07}, an approach very similar to that recommended for 
multiplying dense matrices, to increase the amount of computation -- albeit 
mostly unnecessary -- performed for each global memory access. The outcome is 
that the force computations constitute almost the entire workload. The measured 
speedup of the GPU compared to the CPU is 80x for a relatively small $N_a = 
4096$ SP system. Unfortunately the improvement is misleading, and the approach 
is impractical since the work varies as ${N_a}^2$ rather than $N_a$; even for 
this small size it is over 50x slower than the layer method.

\subsection{Further MD studies}

There are numerous extensions of the basic MD approach for short-range forces 
that cater to different kinds of systems. Two that require relatively 
straightforward modifications for the GPU are discussed below; others requiring 
major changes lie beyond the scope of the present treatment.

(1) Rigid bodies: Modeling molecules based on rigid bodies with multiple 
interaction sites follows directly from the basic method. A nonspherical 
structure can be formed, e.g., from four SP spheres in a rigid tetrahedral 
configuration, spaced to overlap slightly. Spatial orientation can be described 
in different ways, as quaternions or (as used here) rotation matrices 
\cite{mdbk}. Integration of the rotational equations of motion employs a 
generalization of the symplectic leapfrog technique. The additional matrix 
computations during integration and force/torque evaluation are readily adapted 
for the GPU, with neighbor processing and forces still dominating the work. This 
approach was used in a granular segregation study \cite{rap14b}; more complex 
shapes were employed, together with an SP solvent, in modeling molecular
self-assembly \cite{rap18a}.

(2) Polymer chains: Polymers are represented by atoms linked with elastic bonds 
and subject to several interactions. Bond length is governed by the force 
between bonded atoms; bond angle is regulated by a torque that depends on 
adjacent bonds; and twist around the bond is (optionally) governed by a dihedral 
angle that depends on three consecutive bonds. Additionally, there are 
interactions between nonbonded chain atoms, and (where relevant) between chains 
and solvent atoms. These interactions can be reformulated for the GPU so that 
each thread deals with a single atom and all its interactions. In a study of how 
stiff polymers pack into small shells \cite{rap16a}, the length of individual 
chains (8000 atoms) was insufficient for optimal GPU use. However, since 
multiple runs from different initial states were required, a set of independent 
simulations could be carried out simultaneously in a single run -- where chains 
were mutually invisible -- to utilize the GPU effectively; the only practical 
consideration is avoiding excessive cell occupancy, and this was achieved by 
displacing the shell and chain locations for each realization.

\section{Conclusion}

The results of this paper show that MD simulation continues to benefit from 
advances in GPU architecture and performance. These developments, however, do 
have consequences for developers and users: algorithms become more complex and 
GPU efficiency remains far below the theoretical peak; these are necessary 
compromises that do not alter the overall effectiveness of the GPU-based 
approach.

\section*{References}

\bibliographystyle{iopart-num}

\bibliography{gpumd_paper}

\end{document}